\documentclass[twoside,twocolumn,9pt]{article}
\usepackage{extsizes}
\usepackage[super,sort&compress,comma]{natbib} 
\usepackage[version=3]{mhchem}
\usepackage[left=1.5cm, right=1.5cm, top=1.785cm, bottom=2.0cm]{geometry}
\usepackage{balance}
\usepackage{mathptmx}
\usepackage{sectsty}
\usepackage{graphicx}
\usepackage{lastpage}
\usepackage[format=plain,justification=justified,singlelinecheck=false,font={stretch=1.125,small,sf},labelfont=bf,labelsep=space]{caption}
\usepackage{float}
\usepackage{fancyhdr}
\usepackage{fnpos}

\usepackage[english]{babel}
\addto{\captionsenglish}{%
  
}
\usepackage{array}
\usepackage{droidsans}
\usepackage{charter}
\usepackage[T1]{fontenc}
\usepackage[usenames,dvipsnames]{xcolor}
\usepackage{setspace}
\usepackage[compact]{titlesec}
\usepackage{hyperref}

\usepackage{amssymb}
\usepackage{bm}

\usepackage{epstopdf}
\usepackage[varg]{newtxmath}

\definecolor{cream}{RGB}{222,217,201}

\begin{document}

\pagestyle{fancy}
\thispagestyle{plain}
\fancypagestyle{plain}{
\renewcommand{\headrulewidth}{0pt}
}

\makeFNbottom
\makeatletter
\renewcommand\LARGE{\@setfontsize\LARGE{15pt}{17}}
\renewcommand\Large{\@setfontsize\Large{12pt}{14}}
\renewcommand\large{\@setfontsize\large{10pt}{12}}
\renewcommand\footnotesize{\@setfontsize\footnotesize{7pt}{10}}
\makeatother

\renewcommand{\thefootnote}{\fnsymbol{footnote}}
\renewcommand\footnoterule{\vspace*{1pt}%
\color{cream}\hrule width 3.5in height 0.4pt \color{black}\vspace*{5pt}} 
\setcounter{secnumdepth}{5}

\makeatletter 
\renewcommand\@biblabel[1]{#1}            
\renewcommand\@makefntext[1]%
{\noindent\makebox[0pt][r]{\@thefnmark\,}#1}
\makeatother 
\renewcommand{\figurename}{\small{Fig.}~}
\sectionfont{\sffamily\Large\bf}
\subsectionfont{\normalsize}
\subsubsectionfont{\bf}
\setstretch{1.125} 
\setlength{\skip\footins}{0.8cm}
\setlength{\footnotesep}{0.25cm}
\setlength{\jot}{10pt}
\titlespacing*{\section}{0pt}{4pt}{4pt}
\titlespacing*{\subsection}{0pt}{15pt}{1pt}

\fancyfoot{}
\fancyfoot[LO,RE]{\vspace{-7.1pt}\includegraphics[height=9pt]{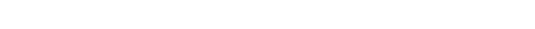}}
\fancyfoot[CO]{\vspace{-7.1pt}\hspace{13.2cm}\includegraphics{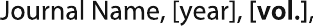}}
\fancyfoot[CE]{\vspace{-7.2pt}\hspace{-14.2cm}\includegraphics{head_foot/RF}}
\fancyfoot[RO]{\footnotesize{\sffamily{1--\pageref{LastPage} ~\textbar  \hspace{2pt}\thepage}}}
\fancyfoot[LE]{\footnotesize{\sffamily{\thepage~\textbar\hspace{3.45cm} 1--\pageref{LastPage}}}}
\fancyhead{}
\renewcommand{\headrulewidth}{0pt} 
\renewcommand{\footrulewidth}{0pt}
\setlength{\arrayrulewidth}{1pt}
\setlength{\columnsep}{6.5mm}
\setlength\bibsep{1pt}

\makeatletter 
\newlength{\figrulesep} 
\setlength{\figrulesep}{0.5\textfloatsep} 

\newcommand{\topfigrule}{\vspace*{-1pt}%
\noindent{\color{cream}\rule[-\figrulesep]{\columnwidth}{1.5pt}} }

\newcommand{\botfigrule}{\vspace*{-2pt}%
\noindent{\color{cream}\rule[\figrulesep]{\columnwidth}{1.5pt}} }

\newcommand{\dblfigrule}{\vspace*{-1pt}%
\noindent{\color{cream}\rule[-\figrulesep]{\textwidth}{1.5pt}} }

\makeatother

\twocolumn[
  \begin{@twocolumnfalse}
\vspace{1em}
\sffamily
\begin{tabular}{m{4.5cm} p{13.5cm} }

\includegraphics{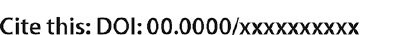} & \noindent\LARGE{\textbf{Impact of Free Energy of Polymers on Polymorphism of Polymer-Grafted Nanoparticles$^\dag$}} \\
\vspace{0.3cm} & \vspace{0.3cm} \\

 & \noindent\large{Masanari Ishiyama,\textit{$^{a}$} Kenji Yasuoka,\textit{$^{a, b}$} and Makoto Asai$^{\ast}$\textit{$^{a, b}$}} \\

\includegraphics{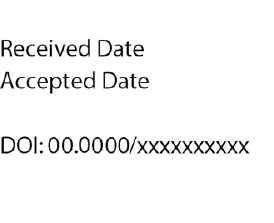} & \noindent\normalsize{Colloidal crystals have gathered wide attention as a model material of photonic crystals, which can control the propagation of light by its crystal structure and lattice spacing, because of the simplicity of the fabrication process and control methods of colloids. However, due to the simple interaction between colloids, the colloidal crystal structure that can be formed is limited. Also, it is difficult to adjust the lattice spacing. Furthermore, colloidal crystals are fragile relatively than other crystals. To solve these problems, we focus on polymer-grafted nanoparticles (PGNP). A PGNP is composed of two different layers: the hard core of a nanoparticle and the soft corona of grafted polymers on the surface. It is predicted that PGNPs with these two distinct layers will have similar behaviors as star polymers and hard spheres. The interaction between PGNPs strongly depend upon their grafting density and the length of the grafted polymer chains, $N$. Thus, PGNP may exhibit polymorphism. Moreover, it is expected that crystals made from PGNPs will be structurally tough due to the entanglement of grafted polymers. The crystal polymorph of PGNP is explored using molecular dynamics simulations. We succeeded in finding FCC/HCP and BCC crystals depending on the length of the grafted polymer chain. When $N$ is small ($N\leqq10$), PGNPs behave like hard spheres. The crystals formed are arranged in  FCC/HCP structure, much like the phase transition observed in an Alder transition. When $N$ is large enough ($N\geqq50$), the increase in the free energy of grafted polymers can no longer be neglected. Thus, the crystals formed in these systems are arranged in BCC structure, which has a lower density than FCC/HCP. When $N$ is not too small or large, FCC/HCP structures are observed when the concentration of PGNPs is low, but a phase transition occurs when the concentration of PGNPs becomes higher with the compression of the system. Again, the increase in free energy of grafted polymers can no longer be neglected and the crystals arrange themselves in a BCC structure. Also, it can be revealed that the lattice spacing of PGNP crystals can be controlled easily and widely by the chain length. These results should play an important role in many future simulations and experimental studies of PGNP crystals.
} \\

\end{tabular}

 \end{@twocolumnfalse} \vspace{0.6cm}

  ]

\renewcommand*\rmdefault{bch}\normalfont\upshape
\rmfamily
\section*{}
\vspace{-1cm}


\footnotetext{\textit{$^{a}$~Department of Mechanical Engineering, Keio University, 3-14-1 Hiyoshi, Kohoku-ku, Yokohama 223-8522, Japan.}}
\footnotetext{\textit{$^{b}$~Keio University Global Research Institute, Keio University, 2-15-45, Mita, Minato-ku, Tokyo 108-8345, Japan. E-mail: m.asai@mech.keio.ac.jp}}




\section{Introduction}
Photonic crystals are spatial periodic structures of refractive index that can eliminate or confine a portion of light with a wavelength comparable to its spatial period from the crystal by Bragg reflection in all directions.\cite{yablonovitch1987inhibited, john1987strong} Furthermore, since the wavelength of light that can be reflected by a photonic crystal depends on the type of crystal and the lattice spacing, it is possible to control the propagation of light by changing the crystal structure. This property is expected to be applied to optical applications such as threshold-free lasers,\cite{noda2000full, akahane2003high, noda2007spontaneous} high-efficiency solar cells,\cite{campbell1987light}, and optical fibers with extremely small allowable bending radius.\cite{birks1995full, birks1997endlessly, russell2003photonic, russell2006photonic} Furthermore, it has been observed that the refractive index of photonic crystals becomes negative at certain frequencies due to anomalous dispersion, in which the refractive index becomes smaller the shorter the wavelength of the incident light.\cite{kosaka1998superprism} This property is expected to be applied to the creation of a meta-material, which is an artificial material to have a property that is not found in naturally occurring materials to electromagnetic waves.\cite{veselago1967electrodynamics, smith2000composite, smith2000negative} As described above, photonic crystals are attracting a great deal of attention as optically important materials.
\par Colloidal photonic crystals, in which colloidal particles are regularly arranged, have attracted much attention due to the simplicity of the fabrication process compared to other types of photonic crystals\cite{jiang2004large, kim2005rapid, wang2006simple} and the fact that the colloids can be directly observed and controlled by laser tweezers.\cite{grier1997optical} However, the crystal structures that can be formed in colloidal crystals are limited due to the simple interaction between the colloids. Furthermore, it is difficult to change the lattice spacing of conventional colloidal crystals. In densely packed colloidal crystals where the particles contact each other, it is necessary to change the size of the particles themselves to change the lattice spacing.\cite{subramania1999optical} In loosely packed (charged) colloidal crystals, where the particles do not contact each other, it is necessary to add ions to the solvent to change the lattice spacing, but it is difficult to precisely adjust the ion concentration in the solvent.\cite{holtz1997polymerized}
\par In this study, we focus on polymer-grafted nanoparticles (PGNP), which are nanoparticles (NP) grafted with polymer chains on their surface.\cite{harton2008mean,kumar2013nanocomposites} PGNP has a soft corona layer and a hard core. It is predicted that PGNPs with these two distinct layers will have similar behaviors as star polymers and hard spheres. When the length of the grafted chains, $N$, is extremely small, PGNPs behave like hard spheres and when $N$ is large enough, PGNPs can be regarded the same as star polymers. Hard spheres are known to exhibit first order phase transition to face-centered cubic (FCC) at a volume fraction of about $49\%$.\cite{alder1957phase} Meanwhile, molecular simulations predicted that star polymers have four different crystal structures: body-centered cubic (BCC), FCC, body-centered orthorhombic (BCO), and diamond structures.\cite{watzlawek1999phase} It is obvious that the difference in the crystal structures of these two types of polymers is induced by the quality of intermolecular interactions. Hence, it is possible that the soft corona layer of the PGNP induces crystal polymorphism. The PGNP behavior found in mid-concentrated system is also interesting. In recent studies of PGNP dynamics by Asai et. Al, it was reported that in areas with small grafting density (large $N$), surface fluctuation makes the PGNP to have a rough surface, and the interlocking of such particles limit the degree of free rotation in a concentrated area where glass transition occurs.\cite{asai2018surface} These interactions between PGNPs affect the shape of grafted polymers as the system is compressed, and thus become critical factors that determine the crystal structure of the PGNPs. Furthermore, crystals formed by PGNPs can be expected to have tough structures due to the entanglement of grafted polymers. The distance between NPs may possibly be freely controllable by adjusting $N$.
\par Furthermore, the existence of crystal polymorphs, including quasicrystals, has been reported in the system with both soft and hard repulsion potential. In a simulation of two-dimensional rigid disks with a staircase-like repulsion potential, which mimics a dendritic micelle with soft alkyls around an aromatic core, a variety of quasicrystals were found.\cite{dotera2014mosaic} And the interaction of particles with a rigid disk at the center and surrounded by soft matter was found to be complicated by the effect of density and the size of the soft region. 
\par Although PGNP has great potential to form crystals that conventional colloidal crystals could not, its theoretical crystal phases have not yet been explored and verified. In this study, we have investigated whether crystal phases can be formed in PGNP by changing the length of the grafted polymer chain using molecular dynamics (MD) simulations.

\section{Simulation Method}
\subsection*{Simulation model}
To simulate grafted polymer chains, we used the coarse-grained bead-spring model of Kremer and Grest (KG) with purely repulsive interactions.\cite{kremer1990dynamics} The beads interacted via the Lennard-Jones (LJ) potential [Eq. (1)].
\begin{align}
    U_{\mathrm{LJ}}(r)&=\left\{
    \begin{array}{ll}
    4\epsilon\left\{\left(\frac{\sigma}{r}\right)^{12}-\left(\frac{\sigma}{r}\right)^{6}\right\} & (r \leqq r_\mathrm{c}) \\
    0 & (r>r_\mathrm{c})
    \end{array}
    \right.
\end{align}
Here, $r$ is the distance between two beads, $\epsilon$ is the interaction strength, and $\sigma$ is the scale unit. We set the cutoff length $r_\mathrm{c}$ of the interaction to $2^{1/6}\,\sigma$ to achieve an excluded volume of chains with purely repulsive interactions in the MD simulations of the coarse-grained KG model. The beads along the chain were connected by an additional unbreakable finitely extensible nonlinear elastic (FENE) potential [Eq. (2)]
\begin{align}
    U_{\mathrm{FENE}}(r)&=\left\{
    \begin{array}{ll}
    -\frac{1}{2}k{R_0}^2\ln\{1-\left(\frac{r}{R_0}\right)^2\} & (r\leqq R_0) \\
    \infty & (r> R_0) 
    \end{array}
    \right.
\end{align}
Here, $k$ is the spring constant and $R_0$ is the maximum bond length. $k=30\, \epsilon/\sigma^2$  and $R_0=1.5\,\sigma$ were used to prevent the chains from crossing over into each other when using the LJ and FENE potentials. We used an expanded LJ interaction for interactions between NP-NP and NP-(polymer bead), expressed as follows:
\begin{align}
    U(r)&=\left\{
    \begin{array}{ll}
    4\epsilon\left\{\left(\frac{\sigma}{r-\Delta}\right)^{12}-\left(\frac{\sigma}{r-\Delta}\right)^{6}\right\} & (r\leqq r_\mathrm{c}+\Delta) \\
    0 & (r>r_\mathrm{c}+\Delta)
    \end{array}
    \right.
\end{align}
$\Delta$ is the shifted distance which ensures that $U(r)$ is 0 when NP-NP or NP-(polymer bead) are in contact. We set $\Delta=4\,\sigma$ and $\Delta=2\,\sigma$ for NP-NP and NP-(polymer bead) interactions, respectively. The diameter of the NP, $2R_\mathrm{n}$, and that of the polymer beads, $2R_\mathrm{b}$ were set to $5.0\,\sigma$ and $1.0\,\sigma$, respectively.

\par
\begin{figure}[t]
\centering
  \includegraphics[width=8.3cm]{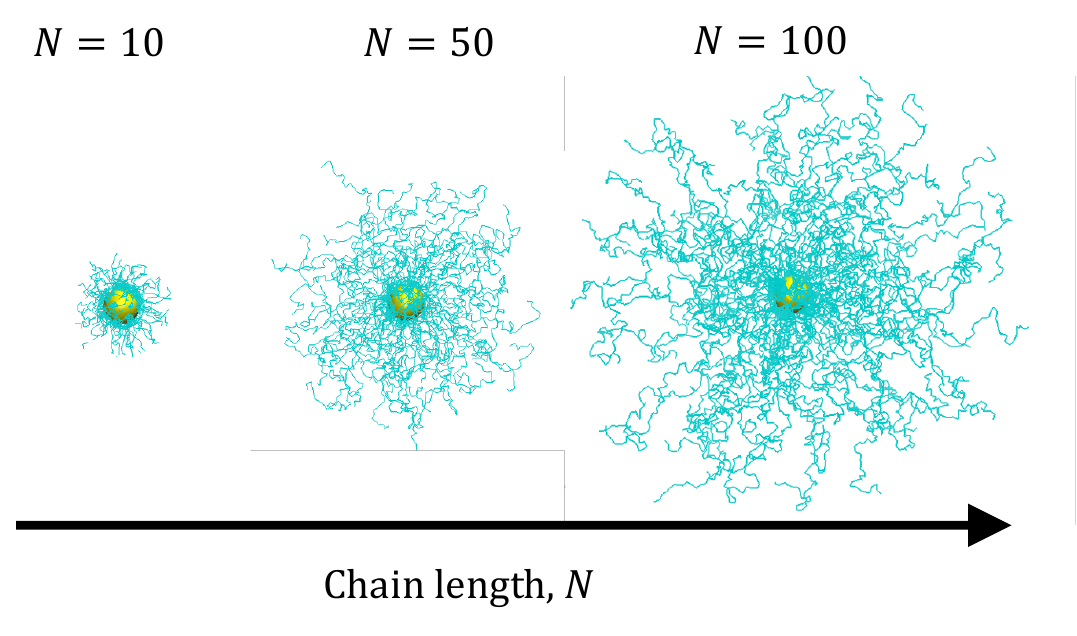}
  \caption{Typical PGNP configurations for $N=10,\,50,\,100$ with the grafting density $\sigma_{\mathrm{g}}=0.89\,\sigma^{-2}$ in dilute solutions in a good-solvent regime. Yellow spheres represent NPs and blue lines represent grafted polymer chains.}
  \label{fgr:PGNP}
\end{figure}

 
In our simulation, PGNP was composed of a NP grafted with $f$ polymer chains on its surface. Here, we fixed $f=70$. The corresponding grafting density, $\sigma_{\mathrm{g}}$, is $0.89\,\sigma^{-2}$. This is the density that can densely fill the surface of the NP with polymer chains. Here, the relationship between the number of grafted polymer chains, $f$, and $\sigma_{\mathrm{g}}$ is expressed as $\sigma_{\mathrm{g}}=f/(4\pi {R_{\mathrm{b}}}^2)$. Grafting points, the points at which the end monomers of polymer chains were fixed to the NP surface, were randomly placed without any overlaps. We examined the cases for systems with the length of the grafted chains $N=10,\,20,\,30,\,50,\,80,\,100$. PGNP with $N=10,\,50,\,100$ are shown in  Fig. \ref{fgr:PGNP}.

\subsection*{MD simulation}
In this study, we used the MD simulation package, LAMMPS.\cite{plimpton1995fast} The dynamics of the polymer beads and NP in our model was described by a Langevin equation with a damping constant $\Gamma=0.5\,m/\tau$ and temperature $T=1.0\,k_\mathrm{B}$, where $k_\mathrm{B}$ is the Boltzmann constant and $m$ is the mass of the monomer. The LJ time scale was given by $\tau=\sigma(m/\epsilon)^{1/2}$. The velocity Verlet algorithm was used to numerically integrate the Langevin equation with a time step $\Delta t=0.005\,\tau$.

\par 256 PGNPs were randomly placed in a simulation box under periodic boundary conditions with volume fraction $\phi=0.0001$. We set 
\begin{align}
    \phi=\frac{4/3\pi({R_\mathrm{n}}^3+fN{R_\mathrm{b}}^3)N_\mathrm{p}}{L^3}
\end{align}
as the volume fraction of PGNPs in this system. Here, $N_\mathrm{p}$ is the number of PGNPs and $L$ is the side length of the cubic simulation box. Next, $NVT$ simulations were performed to equilibrate the system. The system was then slowly enriched by isotropically reducing the size of the simulation box.

\par When $N\leqq80$, the size of simulation box was reduced so that the volume fraction of PGNPs increased by $1\,\%$ for every 50 million time-steps of the calculation run. When $N=100$, simulation was run with 80 million time-steps for increasing the volume fraction of PGNP increased by $1\,\%$. This is because the time steps required for the PGNP to align for phase transition increases since the mass of the PGNP increases as the chain length increases, which reduces the acceleration of the PGNP.

\par For property calculations of the system at the $\phi$ of interest, the simulations were run for 1 million time steps to equilibrate the system, and then another 1 million time steps. At that time, 100 samples of coordinate data of PGNP were obtained for every 1000 time steps, and properties were calculated and averaged.

\section{Results}
\begin{figure}[t]
\centering
  \includegraphics[width=8.3cm]{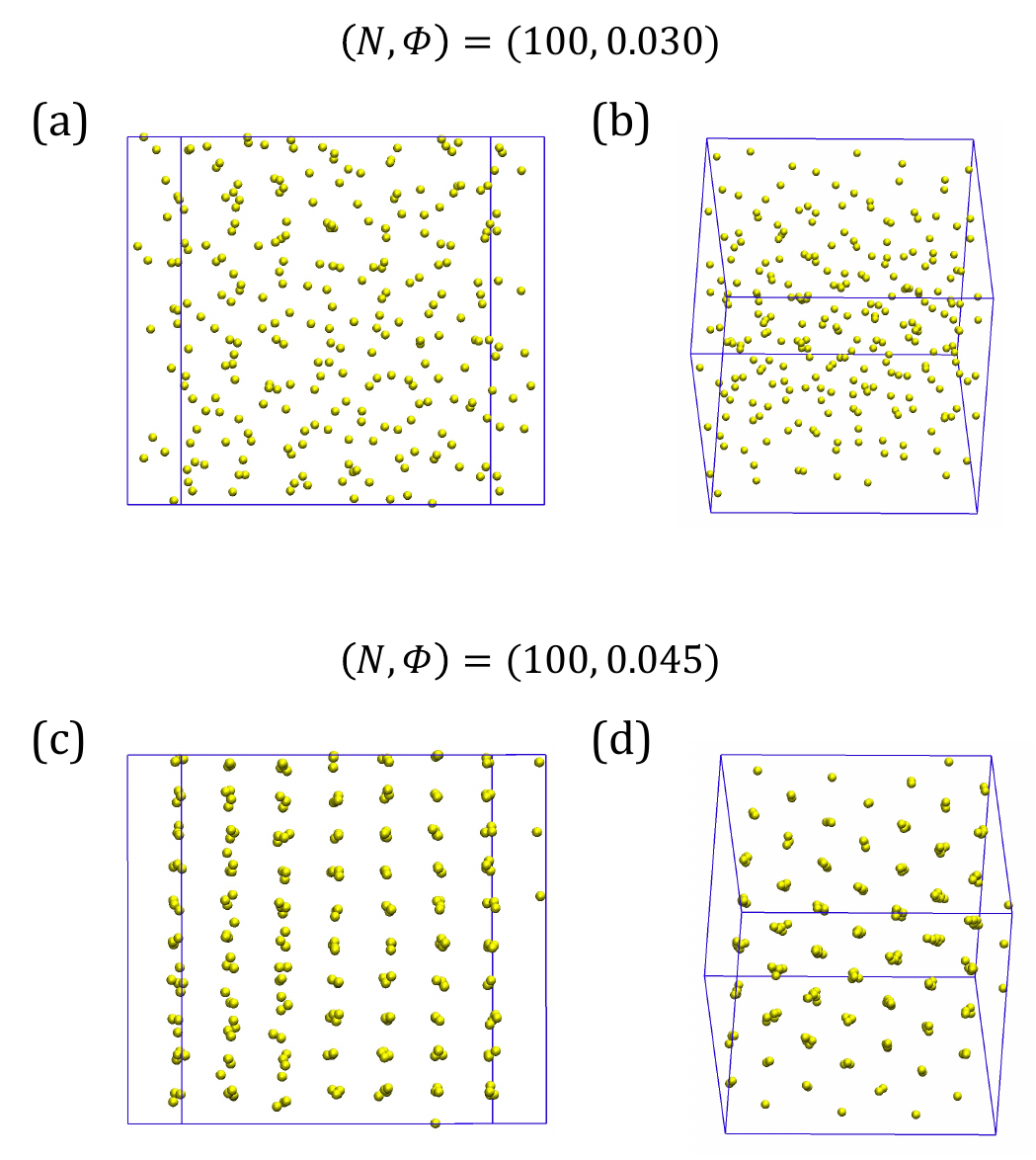}
  \caption{(a, b) Snapshots of a disordered phase of PGNP at $(N, \phi)=(100, 0.030)$ from two different angles. (c, d) Snapshots of a crystal phase of PGNP at $(N, \phi)=(100, 0.045)$ from two different angles. To help with the visualization of the structure, only the NP center has been rendered.}
  \label{fgr:snp100}
\end{figure}

Fig. \ref{fgr:snp100} shows snapshots of the simulation results at $(N, \phi)=(100, 0.030)$ and $(N, \phi)=(100, 0.045)$. PGNPs are disorderly located at $(N,\phi)=(100, 0.030)$, but on the other hand PGNPs are regularly arranged at $(N, \phi)=(100, 0.045)$. From these snapshots, it is confirmed that a phase transition occurred from the disordered phase to the crystal phase in a bulk system of PGNP for $N=100$. For $N=10, 20,30,50,80$, crystallization also occurred as shown in the Supporting Information (Section A).

\par
\begin{figure}[t]
\centering
  \includegraphics[width=8.3cm]{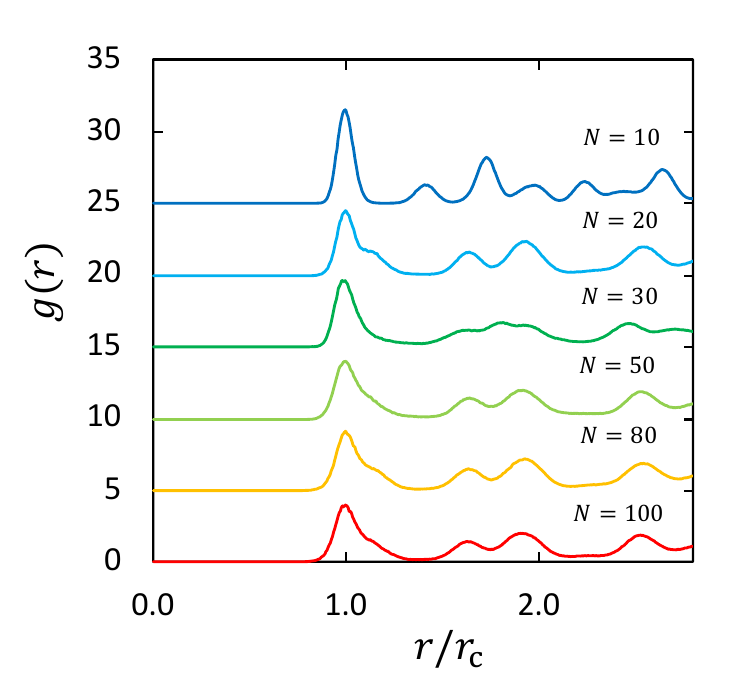}
  \caption{Radial distribution function, $g(r)$, between the NP centers in the crystal phase. The distance $r$ is scaled by $r_{\mathrm{c}}$, the distance at the first peak of $g(r)$.}
  \label{fgr:rdf}
\end{figure}

 The radial distribution function of the NP centers, $g(r)$, in the crystal phase is presented in Fig. \ref{fgr:rdf} for each $N$. All $g(r)$ has sharp peaks which clearly indicate the presence of a long-range ordered structure. Significantly different peaks are observed when $N=10$ and $N\geqq20$, indicating that different ordered structures are formed. From some of these distinctive peak positions, it could be predicted that when $N=10$, the crystals have either FCC or hexagonal close-packed (HCP) structure and when $N\geqq20$, the crystals have a BCC structure. Moreover, when $N\geqq20$, the distance between peaks of $g(r)$ are relatively the same for each $N$, while peak positions seem to shift to the right as $N$ becomes bigger. In other words, the greater $N$ is, the greater the distinctive length of the ordered structure.

\par In order to analyze the structure of PGNP in more detail, we calculated the averaged local bond order parameters of the NP, $\overline{q}_4$, $\overline{q}_6$,and $\overline{w}_4$, which have developed by Lechner and Dellago.\cite{lechner2008accurate} Averaged local bond order parameters are a modified version of local bond order parameters based on spherical harmonics, also known as Steinhardt order parameters which are often used to determine crystal structures in molecular simulations.\cite{steinhardt1983bond} This modified method considerably improves the accuracy with which different crystal structures can be distinguished. The order parameters $\overline{q}_l(i)$, $\overline{w}_l(i)$ are defined as:
\begin{align}
    \overline{q}_l(i)&= \sqrt{\frac{4\pi}{2l+1}\sum_{m=-l}^{l}|\overline{q}_{lm}(i)|^2}\\
    \overline{w}_i(i)&=\frac{\sum_{m_1+m_2+m_3=0}
	\begin{pmatrix} l&l&l&\\m_1&m_2&m_3\end{pmatrix}
    \overline{q}_{lm_1}(i)\overline{q}_{lm_2}(i)\overline{q}_{lm_3}(i)}
    {(\sum_{m=-l}^{l}|\overline{q}_{lm}(i)|^2)^{\frac{3}{2}}}
\end{align}
where
\begin{align}
    \overline{q}_{lm}(i)&=\frac{1}{\tilde{N}_b(i)}\sum_{k=1}^{\tilde{N}_b(i)}q_{lm}(k)\\
    q_{lm}(i)&=\frac{1}{N_b(i)}\sum_{j=1}^{N_b(i)}Y_{lm}(\bm{r}_{ij})
\end{align}
Here, $\overline{q}_{lm}(i)$ is the average of $q_{lm}(i)$ by $\tilde{N}_b(i)$. $q_{lm}(i)$ is the complex vector of particle $i$ and $\tilde{N}_b(i)$ is the number of nearest neighbors of particle $i$ plus the particle $i$ itself. $l$ is a free integer parameter, and $m$ is an integer that runs from $m=-l$ to $m= +l$. The integers $m_1$, $m_2$, and $m_3$ run from $-l$ to $+l$, but only combinations with $m_1+m_2+m_3=0$ are allowed. The term in parentheses is the Wigner $3-j$ symbol.\cite{landau2013quantum} $N_b(i)$  is the number of nearest neighbors of particle $i$, the functions $Y_{lm}(\bm{r}_{ij} )$ are the spherical harmonics and $\bm{r}_{ij}$ is the vector from particle $i$ to particle $j$. When examining liquid-solid transitions, $\overline{q}_4-\overline{q}_6$ plane and $\overline{q}_4-\overline{w}_4$ plane are most commonly chosen as they are useful to identify BCC, FCC and HCP structures.

\par
\begin{figure*}
\centering
  \includegraphics[width=17.1cm]{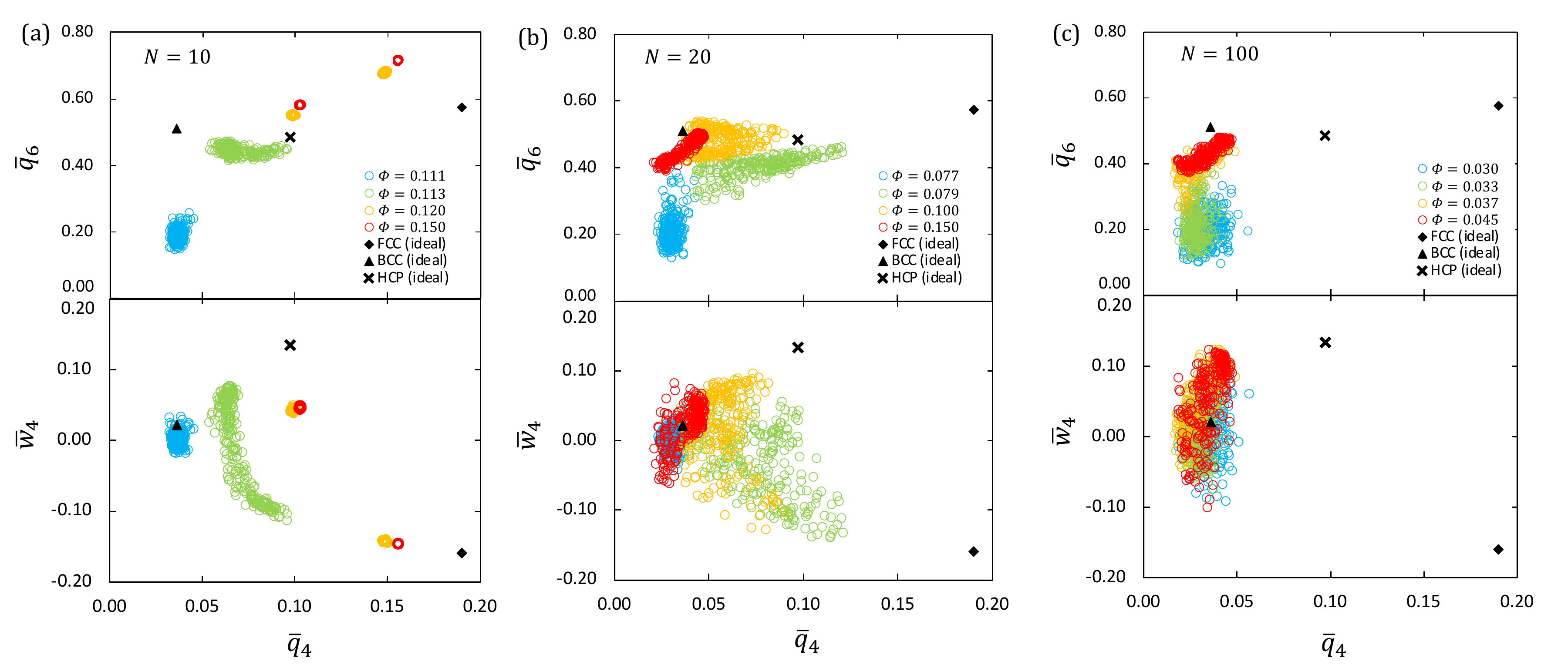}
  \caption{Plots of $\overline{q}_6$ ($\mathbf{top}$) and $\overline{w}_4$ ($\mathbf{bottom}$) versus $\overline{q}_4$ for (a) $N=10$,(b) $N=20$ and (c) $N=100$. Points of all $256$ PGNPs at each structure are shown. The volume fraction of each point is denoted in the legend. For comparison, data points for perfect FCC, BCC and HCP crystals are also included.}
  \label{fgr:bop_ex}
\end{figure*}

 As an example, the $\overline{q}_4-\overline{q}_6$ and $\overline{q}_4-\overline{w}_4$ plane of order parameters for $N=10$, $20$ and $100$ are shown in Fig. \ref{fgr:bop_ex}. For each sample, the order parameters were calculated for 256 NPs and plotted. Notice that when $N=10$, $\overline{q}_6$ and $\overline{q}_4$ of the order parameter are small in lower density region ($\phi=0.111$). This indicates that the NPs are in a disordered state. However, when the density becomes higher ($\phi=0.113$), there is a drastic change. The plot distribution in the $\overline{q}_4-\overline{q}_6$ plane seems to span over HCP and BCC regions while the plot distribution in the $\overline{q}_4-\overline{w}_4$ plane seems to span only over the FCC region. When the system is further compressed, the distribution distinctively splits in two phases (FCC and HCP) when $\phi=0.120$. In other words, when $N=10$, three polymorphs (FCC/HCP/BCC) co-exist shortly after crystallization, but as the system is compressed, BCC structures cannot grow and disappears. Ultimately, only two polymorphs (FCC/HCP) co-exist. This result agrees well with the radial distribution function shown in Fig. \ref{fgr:rdf}.

Now take a look at when $N=20$. When $\phi=0.077$, NPs are in a disordered state. However, the state drastically changes when $\phi=0.079$ and three polymorphs (FCC/HCP/BCC) co-exist as it did when $N=10$. When the system is further compressed, a phase transition to BCC was observed at $\phi=0.150$. The same tendency was observed when $N=30$.

Now take a look at when $N=100$ (the same tendency was also observed when $N=50$ and $80$). Notice that the phase drastically changes from a disordered state to BCC when $\phi=0.037$. The distributions of order parameters for each PGNP for other $N$ are shown in the Supporting Information (Section B). 

\par
\begin{figure*}
\centering
  \includegraphics[height=6cm]{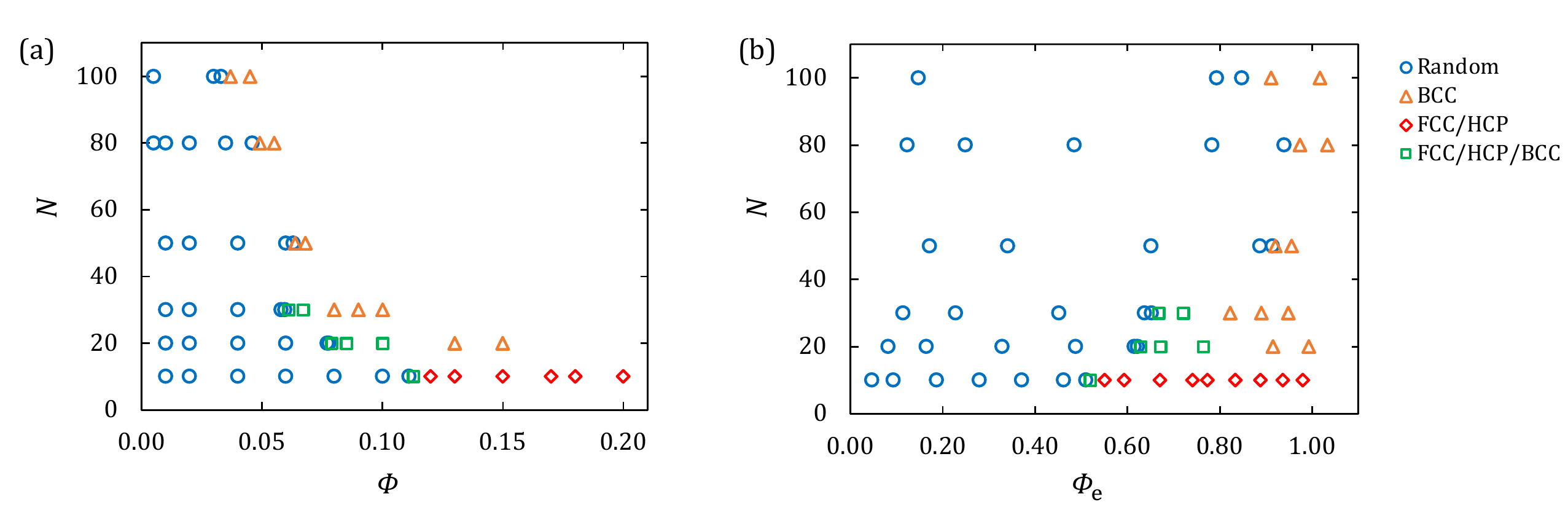}
  \caption{(a) Phase diagram plots of $N$ versus $\phi$ of PGNP. (b) Phase diagram plots of $N$ versus $\phi_\mathrm{e}$ of PGNP. The phase of each point is denoted in the legend.}
  \label{fgr:diagram}
\end{figure*}

 The transition phase diagram for the $\phi-N$ plane of PGNPs is shown in Fig. \ref{fgr:diagram}. The volume fraction of the system, $\phi_\mathrm{e}$, is redefined by approximating PGNP as an isotropic spherical particle. Here, the effective radius of PGNPs consists of the radius of NP ($R_\mathrm{n}$) and the thickness of the corona layer made from the grafted polymers, which is based on the end-to-end distance ($R_\mathrm{e}$) of the grafted polymers. The effective radius is defined as follows. 
\begin{align}
    \phi_{\mathrm{e}}=\frac{4/3\pi {(R_{\mathrm{n}}+\sqrt{<{R_{\mathrm{e}}}^2>})}^3 N_{\mathrm{p}}}{L^3}
\end{align}

Let’s first take a look when $N=10$. The phase shifts to FCC/HCP/BCC at a concentration of $\phi=0.113$, which is equivalent to $\phi_\mathrm{e}=0.520$. The phase then shifts to FCC/HCP at a concentration of $\phi=0.120$, which is equivalent to $\phi_\mathrm{e}=0.551$. This is almost equivalent to the volume fraction of Alder transition in hard spheres ($0.490-0.500$).\cite{alder1957phase,dolbnya2005coexistence} The difference with this transition and the Alder transition of hard spheres is that BCC can also be observed in the initial crystallization phase. This is likely due to particles choosing to form BCC structures, which have lower density, in order to avoid having the corona layer get compressed in locally concentrated areas. In other words, it is likely that both density and entropy of grafted polymer chains are trying to maximize, and thus the three polymorphs co-exist in this concentration region. As the system is further compressed, PGNPs start behaving like hard spheres because the length of grafted polymer chains is short. It is assumed that BCC structures disappear and the phase transits to FCC/HCP as a result. 

When $N=20$ and $30$, the phase shifts to FCC/HCP/BCC when the concentration is low (for $N=20$, $\phi=0.079$ and $\phi_\mathrm{e}=0.629$ ) (for $N=30$, $\phi=0.061$ and $\phi_\mathrm{e}=0.667$ ) as after-mentioned. This is likely due to the same reason as was seen when $N=10$. Note, however, that the area in which the three phases co-exist is wider than when $N=10$. When $N$ becomes greater, the corona layer becomes softer and thus easier to compress. Unlike the case when $N=10$, FCC/HCP area disappears when the system is further compressed, and BCC becomes dominant. It is assumed that in systems with larger $N$, the urge to maximize the entropy of grafted polymer chains becomes dominant, and the particles choose to take BCC structures, which have lower density than FCC/HCP. 

When $N\geqq50$, the phase directly shifts to BCC from a disordered state. When $N$ is large enough, PGNPs do not behave like hard spheres. Rather, they behave like soft particles. Phase transition is observed at a concentration of $\phi_\mathrm{e}=0.900$. It seems that this does not depend on the size of $N$.

\par
\begin{figure*}
\centering
  \includegraphics[height=6cm]{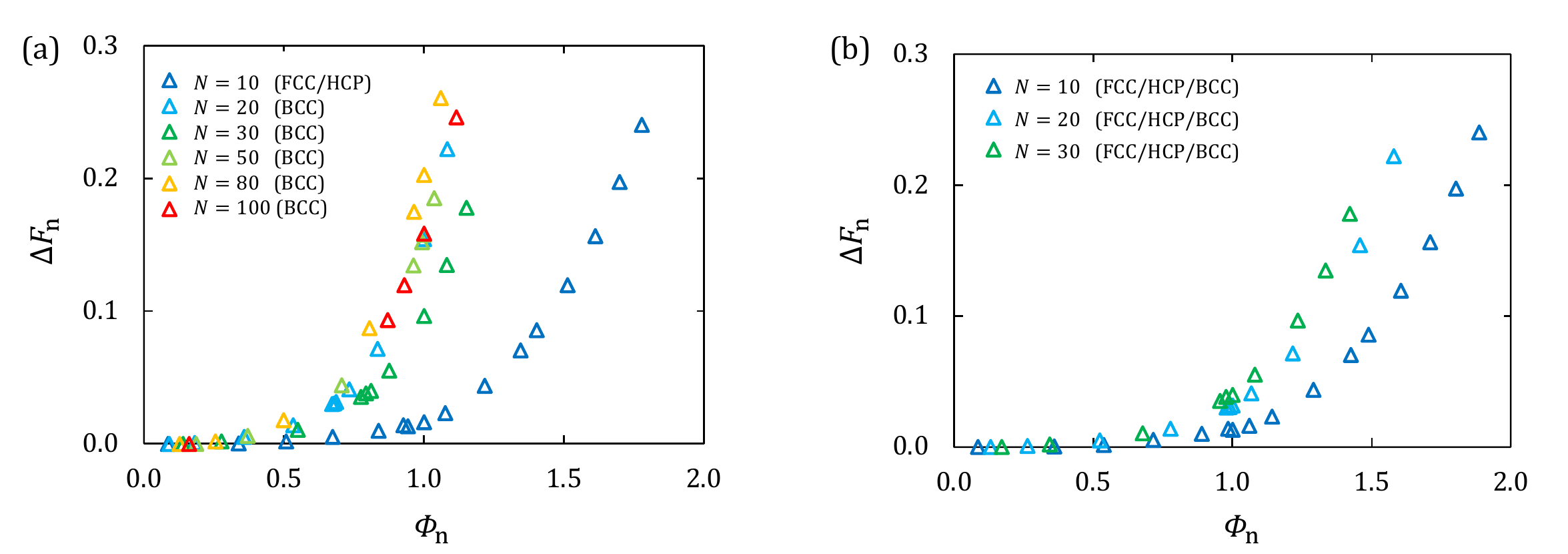}
  \caption{Normalized free-energy difference, $\Delta F_{\mathrm{n}}$, versus $\phi_{\mathrm{n}}$ for $N=10, 20, 30, 50, 80, 100$. The volume fraction $\phi_{\mathrm{n}}$ is scaled to be 1 at the crystallization for comparison.}
  \label{fgr:delta-f}
\end{figure*}

 TIn this section, we will discuss free energies of PGNPs in aim to understand the phase diagram of $\phi-N$ plane in detail. The chain free energy of PGNP can be calculated by employing the Flory-type arguments as follows\cite{likos2006soft}:
\begin{align}
    \frac{F(R)}{k_{\mathrm{B}}T}=\frac{3fR^2}{2Nb^2}+\frac{\nu f^2N^2}{2R^3}+\frac{wb^6N^3f^3}{R^6}
\end{align}
Here, $R$ is the chain size and $b$ is Kuhn length. We used the root mean squared radius of gyration of chains, $\sqrt{<{R_{\mathrm{g}}}^2>}$, as a chain size. $\nu$ is the Flory exponent and $w$ is the ternary-contact excluded volume parameter which typically is of order unity. In good solvent conditions the Flory exponent term dominates over the ternary term and the latter can be ignored. From the log-log plots $N$ versus the radius of gyration of chains, $R_{\mathrm{g}_0}$, in dilute a system, we got the regression line $R_{\mathrm{g}}=0.408N^{0.674}$ and we adopted $b=0.408$ and $\nu=0.674$ (see Supporting Information, Section C). We calculated free-energy difference, $\Delta F_{\mathrm{n}}=(F(R_{\mathrm{g}})-F_0(R_{\mathrm{g}_0})/F_0(R_{\mathrm{g}_0})$. The results are shown in Fig. \ref{fgr:delta-f}.

Fig. \ref{fgr:delta-f} (a) shows a comparison of $\Delta F_{\mathrm{n}}$ for $N=10, 20$, and $30$. Concentrations are normalized with the concentration at which the phase shifts to the co-existing state of the three phases (FCC/HCP/BCC). Although we can observe the tendency of the fluctuation getting bigger with the increase in $N$, $\Delta F_{\mathrm{n}}$ does not change much until the phase transition point. Thus, it could be said that the decrease in grafted polymer entropy has little impact to the phase transition and that PGNPs generally behave like hard spheres in the co-existing state. 

Fig. \ref{fgr:delta-f} (b) shows the comparison of $\Delta F_{\mathrm{n}}$ in the range of $N=20$ to $N=100$. Concentration in this figure was normalized with the concentration at which the phase shifted to BCC. As a result, a master curve that does not depend on $N$ was drawn, indicating that the increase in free energy of grafted polymers is the direct cause for phase transition to BCC. We also confirmed that the phase transition to BCC was triggered when $\Delta F_{\mathrm{n}}\sim0.15$, regardless of the size of $N$.

\par
\begin{figure}[t]
\centering
  \includegraphics[width=8.3cm]{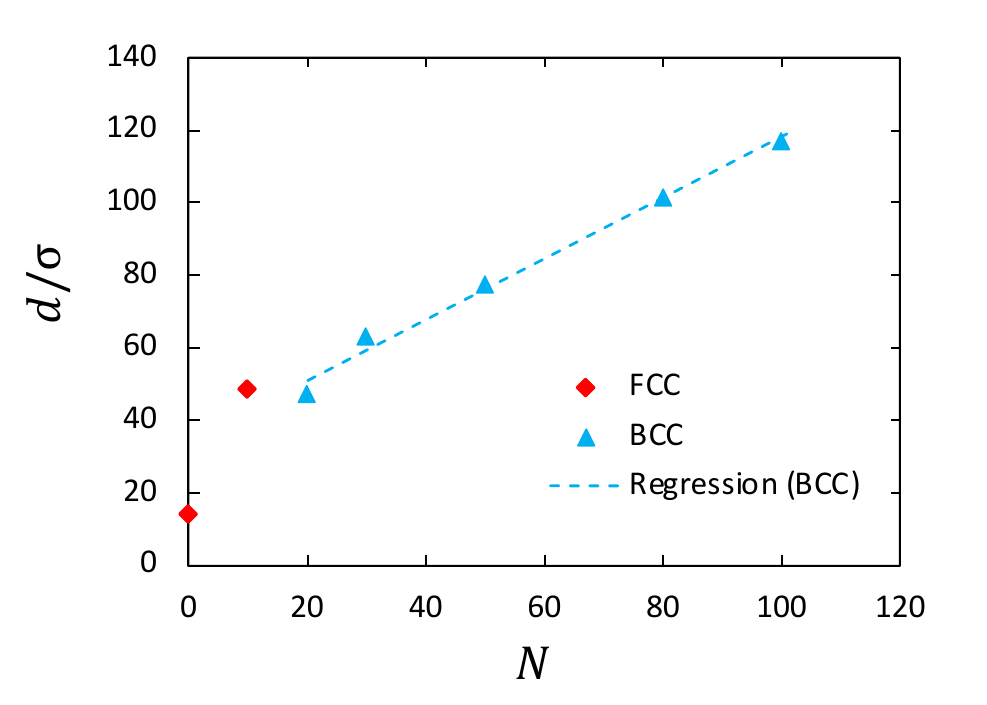}
  \caption{Plots of lattice constant, $d$, of PGNP crystals versus $N$ of PGNP. The type of crystal is denoted in the legend. The coefficient of determination, $R^2$, is 0.9898.}
  \label{fgr:lattice}
\end{figure}

 Plots of lattice constant, $d$, of BCC crystals of PGNP versus $N$ are shown in Fig. \ref{fgr:lattice}. For $N=20, 50, 80, 100$, the relationship between $d$ and $N$ is clearly linear, and the regression line is as follows:
\begin{align}
    d=0.418N+17.2 \label{d-N}
\end{align}
By using Eq. \eqref{d-N} as an indicator when synthesizing PGNP, we can adjust the lattice spacing of the crystals and obtain the desired PGNP crystals under the current calculation conditions ($\sigma_{\mathrm{g}}=0.89\,\sigma^{-2}$, $R_\mathrm{n}=2.5\,\sigma$ and $R_\mathrm{b}=0.5\,\sigma$). Adjusting the crystal lattice spacing by the chain length is an unconventional concept. And since the adjustment of the lattice spacing by this new parameter can be easily done during the synthesis of PGNP, PGNP can be expected to be even more promising as an optical material. Considering the size ratio of NP and monomer beads, we should be able to find a relationship between $d$ and $N$ for PGNP in general conditions.

\par In the previous study of star polymers, the factors of phase transition of star polymers are discussed as follows\cite{likos2006soft}: for high grafting density (strong screeing), packing effect dominate and stabilze the FCC, whereas for low grafting density, energy plays an important role stabilizing the BCC structure. This means that the crystal structure of star polymers depends on the softness of the polymer layer. The larger the grafting density, the harder the star polymer becomes, resulting in FCC, and the smaller the grafting density, the softer the star polymer becomes, resulting in BCC. Similarly, it was found that the crystal structure of PGNP also depends on the softness of polymer layer. For high grafting density, the shorter the chains, the harder the PGNP becomes, resulting in FCC, and the longer the chains, the softer the PGNP becomes, resulting in BCC. However, it should be noted that they behave differently at high density. Star polymers with very long chains, which become FCC, are strongly affected by polymers and change to BCO and diamond at high density. On the other hand, PGNP for $N=10$ behave like hard sphere colloid and remains FCC even at high density. In other words, PGNP for $N=10$ is less affected by polymers due to their short chains even at high density, and they are solid in both dilute and concentrated systems. This may be the reason why PGNP for $N=10$ does not change from FCC to BCC even when they acquire enough free energy change to change to BCC as shown in Fig. \ref{fgr:delta-f}. As an extension of previous work on star polymer crystals, we have succeeded to elucidate the effect of chain length on the crystal phase transition. However, in mapping the phase diagram of star polymers to that of PGNP, it is important to note the limit of grafting density of PGNP which depends on the size ratio of NP to bead, i.e. the curvature of NP. In this study, the maximum number of chains that could be grafted onto a single NP with curvature $1/R_\mathrm{n}=2/5$ was $70$. But, if NP have larger diameter, the curvature will be even smaller. Then, the chains can fill the NP surface more and the polymer layer will be harder, resulting in FCC. In fact, in the phase diagram of star polymers, FCC is stabilized when the number of grafts is about $90$ or more\cite{likos2006soft}. Therefore, mapping the phase diagram of star polymers to the phase diagram of PGNP, the diamond structure of PGNP obtained by further compression FCC crystals is expected to appear at very high grafting density ($f>90$ or $\sigma_\mathrm{g}>1.15$) and longer chain length ($N>50$). Due to its extremely high refractive index, the diamond structure of PGNP is very important for optical applications and is expected to be discovered. From the above, considering the two parameters of chain length and graft density, we can say that there are two types of factors that lead to FCC: short chains and very high grafting density. We have succeeded in linking the predicted phase diagram of star polymers with the one of PGNPs.

\section{Conclusions}
In this study, we conducted MD simulations to explore the bulk state of PGNPs, especially its crystal structures. We focused on the effect that $N$ of grafted polymers had on the crystal phases and confirmed that the bulk state of PGNPs had either FCC, HCP, or BCC phase.

By approximating PGNPs as isotropic spheres, we were able to draw a phase diagram that included the extreme state of $N=0$ (when PGNPs behaved as hard spheres). Furthermore, we were able to uniformly discuss the effect that grafted polymers had on crystal phase transitions in terms of free energy of the grafted polymer chains. It is expected that these discussions will provide important insight to further explore crystal layers yet to be discovered in PGNPs. It was also confirmed that the lattice space of PGNPs crystals can be easily and systematically controlled by $N$. 

By investigating the free-energy difference of PGNP chains, we found a master curve in $\Delta F_{\mathrm{n}}$ versus $\phi_{\mathrm{n}}$ that explains the phase transition to BCC crystals. This means that we have succeeded in extracting the amount of free energy change required for the phase transition to BCC crystals as a universal quantity. And similar to the phase transition of star polymer, we were able to show that the phase transition of PGNP to BCC occurs when the corona layer is soft because the interaction energy becomes dominant, and the phase transition to FCC occurs when the corona layer is hard because the packing effect becomes dominant. However, star polymers became hard and changed to FCC crystals when the chains were very long and the grafting density was extremely high, whereas PGNP became hard and changed to FCC crystals when the chains were short and the grafting density was relatively high. The reason why FCC crystals did not appear in PGNP with long chains is thought to be that the curvature of the NP surface in this study was so large that the grafting density could not be made extremely high, and the chains were not strongly affected by each other's excluded volume effect and did not behave like a hard brush sufficiently. Based on the above, mapping the phase diagram of star polymer to that of PGNP, the diamond structure of PGNP with very high refractive index is expected to appear at very high grafting density ($f>90$ or $\sigma_{\mathrm{g}}>1.15$) and longer chain length ($N\geqq50$). These results should play an important role in many future simulation and experimental studies of PGNP crystals.

\section*{Conflicts of interest}
There are no conflicts to declare.

\section*{Acknowledgements}
The computation was carried out using the computer resource offered under the category of General Projects by Research Institute for Information Technology, Kyushu University.



\balance


\bibliography{rsc} 
\bibliographystyle{rsc} 

\end{document}